\documentclass[12pt]{article}
\usepackage{graphicx}

\usepackage{maybemath}

\usepackage[italic]{hepparticles}

\usepackage{heppennames}

\newcommand\pubnumber{SNSN-323-63}
\newcommand\pubdate{\today}

\def\NotreDame{
University of Notre Dame Du Lac, South Bend, IN 46556, USA}

\def\Title#1{\begin{center} {\Large #1 } \end{center}}
\def\Author#1{\begin{center}{ \sc #1} \end{center}}
\def\Address#1{\begin{center}{ \it #1} \end{center}}

\newcommand\pubblock{\rightline{\begin{tabular}{l} \pubnumber\\
         \pubdate  \end{tabular}}}
\newenvironment{Abstract}{\begin{quotation}  }{\end{quotation}}
\newenvironment{Presented}{\begin{quotation} \begin{center} 
             PRESENTED AT\end{center}\bigskip 
      \begin{center}\begin{large}}{\end{large}\end{center} \end{quotation}}
\def\Acknowledgements{\bigskip  \bigskip \begin{center} \begin{large}
             \bf ACKNOWLEDGEMENTS \end{large}\end{center}}

\def\beq{\begin{equation}}
\def\eeq#1{\label{#1}\end{equation}}
\def\eeqn{\end{equation}}

\def\beqa{\begin{eqnarray}}
\def\eeqa#1{\label{#1}\end{eqnarray}}
\def\eeqan{\end{eqnarray}}

\let\bar=\overbar

\def\eg{{\it e.g.}}

\def\Dslash{\not{\hbox{\kern-4pt $D$}}}
\def\dslash{\not{\hbox{\kern-2pt $\del$}}}

\def\BR{\mbox{\rm BR}}

\def\msb{{\bar{\ssstyle M \kern -1pt S}}}

\def\BB         {\ensuremath{\PB{}\PaB}\xspace}

\def\epem       {\ensuremath{\Pep\Pem}\xspace}

 \def\eg         {\ensuremath{E_{\gamma} }\xspace}

 \def\egcms      {\ensuremath{E^{*}_{\gamma}}\xspace}

 \def\mes        {\ensuremath {M_\mathit{ES}}\xspace}
 
 \def\de         {\ensuremath {\Delta E^{*}}\xspace}

 \def\mupsq      {\ensuremath{\mu_{\pi}^2}\xspace}
 \def\mb         {\ensuremath{m_{b} }\xspace}

 \def\acp        {\ensuremath{A_{CP}}\xspace}
 
 \def\amcp       {\ensuremath{A^{\mathrm{meas}}_{CP}}\xspace}

 \def\bsg        {\ensuremath{\Pqb\to\Pqs\Pgg}}
 \def\bdg        {\ensuremath{\Pqb\to\Pqd\Pgg}\xspace}

 \def\bxsg       {\ensuremath{\PB \to X_{s} \Pgg}\xspace}
 
 \def\bxsdg       {\ensuremath{\PB \to X_{s+d} \Pgg}\xspace}
 
 \def\bxclnu      {\ensuremath{\PB \to X_{c} \ell \nu }\xspace}
 \def\bxulnu      {\ensuremath{\PB \to X_{u} \ell \nu }\xspace}

\def\aveDelta#1 {\ensuremath{\langle \Delta_{total}#1 \rangle}\xspace}

\def\valerr#1#2#3 {\ensuremath{{#1}^{+#2}_{-#3}}\xspace}

\usepackage{relsize}
\def\babar{\mbox{\slshape B\kern-0.1em{\smaller A}\kern-0.1em
    B\kern-0.1em{\smaller A\kern-0.2em R}}}

\def\mes        {\mbox{$m_{\rm ES}$}\xspace}

\newcommand{\tev}{\ensuremath{\mathrm{\,Te\kern -0.1em V}}\xspace}
\newcommand{\gev}{\ensuremath{\mathrm{\,Ge\kern -0.1em V}}\xspace}
\newcommand{\mev}{\ensuremath{\mathrm{\,Me\kern -0.1em V}}\xspace}
\newcommand{\kev}{\ensuremath{\mathrm{\,ke\kern -0.1em V}}\xspace}
\newcommand{\ev}{\ensuremath{\mathrm{\,e\kern -0.1em V}}\xspace}
\newcommand{\gevc}{\ensuremath{{\mathrm{\,Ge\kern -0.1em V\!/}c}}\xspace}
\newcommand{\mevc}{\ensuremath{{\mathrm{\,Me\kern -0.1em V\!/}c}}\xspace}
\newcommand{\gevcc}{\ensuremath{{\mathrm{\,Ge\kern -0.1em V\!/}c^2}}\xspace}
\newcommand{\mevcc}{\ensuremath{{\mathrm{\,Me\kern -0.1em V\!/}c^2}}\xspace}

\def\mus  {\ensuremath{\rm \,\mus}\xspace}

\def\mus        {\ensuremath{\,\mu{\rm s}}\xspace}    

\def\pipi  {\ensuremath{\pi^+\pi^-}\xspace}
\def\pim   {\ensuremath{\pi^-}\xspace}
\def\pip   {\ensuremath{\pi^+}\xspace}
\def\piz   {\ensuremath{\pi^0}\xspace}
\def\Kp    {\ensuremath{K^+}\xspace}

\def\CP                {\ensuremath{C\!P}\xspace}
\def\btosgam {\ensuremath {b\to s\gamma} }
\def\btodgam {\ensuremath {b\to d\gamma} }
\def\BtoXsgam  {\ensuremath{B\to X_s\gamma}}
\def\BtoXdgam  {\ensuremath{B\to X_d\gamma}}

\begin{document}
\begin{titlepage}
\pubblock

\vfill
\Title{$b\rightarrow s\gamma$ and $b\rightarrow d\gamma$ (B factories)}
\vfill
\Author{ Wenfeng Wang} 
\Address{\NotreDame}
\vfill
\begin{Abstract}

The photon spectrum in $B\rightarrow X_{s,d} \gamma$ decay, where $X_s(d)$ is any
strange (non-strange) hadronic state, is studied using data samples of
$e^+e^- \rightarrow \Upsilon(4S) \to \PB\PaB$ decays collected by the \babar\ and
 Belle experiments. Here I present the latest measurements of the branching fraction and spectral moments from $B\to X_s \gamma$  decays by Belle
and the direct $\CP$ asymmetry $\acp(B \to X_{s+d}\gamma)$ measured at \babar. The determination of $|V_{td}/V_{ts}|^2$ is also presented.

\end{Abstract}
\vfill
\begin{Presented}
6th International Workshop on the CKM Unitarity Triangle(CKM2010),
\end{Presented}
\vfill
\end{titlepage}
\def\thefootnote{\fnsymbol{footnote}}
\setcounter{footnote}{0}

\section{Introduction}


 The electromagnetic radiative process $b\rightarrow q\gamma$ ($q=s,d$) proceeds at leading order via the loop diagram in the Standard Model (SM). 
Here the SM predication of the inclusive rate $\Gamma(\bxsg)$ can be equated with
the precisely calculable partonic rate $\Gamma(\bsg)$ at the level of a few percent~\cite{Bigi:1992su} (heavy quark duality). An 
extraordinary theoretical effort has led to a precision SM prediction for the branching fraction at the next-to-next-to-leading order (four-loop), $\BR(\bxsg) = (3.15 \pm 0.23) \times 10^{-4}$\,($\eg > 1.6\gev$)~\cite{Misiak:2006zs}, where \eg is the photon energy measured in the rest frame of the \PB meson. 
The possibility for new heavy particles to enter into the loop at leading order could cause significant deviations from the SM prediction. A recent review can be seen in ~\cite{Hurth:2010}.

%

New physics can also significantly enhance the direct \CP asymmetry for \bsg\ 
and \bdg decay~\cite{Hurth:2005CP} without changing the branching fraction. 
We define
\begin{equation}
  \acp = \frac{\Gamma(\bsg + \bdg)-\Gamma(\Paqb\to\Paqs\Pgg + \Paqb\to\Paqd\Pgg)}
       {\Gamma(\bsg + \bdg)+\Gamma(\Paqb\to\Paqd\Pgg + \Paqb\to\Paqd\Pgg)}
  \label{eq:ACPdefinition}
\end{equation}
which is $\sim 10^{-6}$ in the SM, with nearly exact cancellation
of opposite asymmetries for \bsg\ and \bdg.  
Thus any non-zero measurements of this joint asymmetry is an indication of new physics. 

The shape of the  photon energy spectrum, which is insensitive 
to non-SM physics~\cite{Kagan:1998ym}, can be used to determine the Heavy Quark Expansion (HQE)parameters, \mb and \mupsq, related 
to the mass and momentum of the  \Pqb quark within the \PB  meson. These parameters can be used to 
reduce the error in the extraction of the CKM matrix elements $V_{cb}$ and $V_{ub}$ from the inclusive 
semi-leptonic  \PB-meson decays, \bxclnu and \bxulnu ($\ell = e$ or $\mu$)~\cite{VcbVub} .

The inclusive rate for $b\rightarrow d\gamma$\ is suppressed compared to $b\rightarrow s\gamma$ by a factor 
$|V_{td}/V_{ts}|^2$ in the SM. This ratio can also be obtained from the $B_d$ and $B_s$ 
mixing frequencies~\cite{bsmixing}. 
New physics effects would enter in different ways in mixing and radiative decays.  
Measurements of $|V_{td}/V_{ts}|$ using the exclusive modes 
$B\rightarrow (\rho,\omega)\gamma$ and $B\to K^*\gamma$~\cite{bellerhog,babarrhog} are now 
well-established, with  
theoretical uncertainties of 7\%~\cite{BJZ}.
A measurement of inclusive $b\rightarrow d\gamma$ relative to $b\rightarrow s\gamma$ would  
determine $|V_{td}/V_{ts}|$ with reduced theoretical uncertainties.  
We parametrize the inclusive ratio (following~\cite{AAG}) by:
\begin{equation}
{{\BR(b\rightarrow d\gamma)}\over{\BR(b\rightarrow s \gamma)}} = 
\zeta^2\left|{{V_{td}}\over{V_{ts}}}\right|^2 (1+\Delta R)
\end{equation}
where $\zeta$ accounts for any remaining SU(3) breaking and 
$\Delta R$ accounts for weak annihilation in $B^+$ decays.


B factories, \babar~\cite{Babar:NIM} and Belle~\cite{Belle:NIM}, already accumulated more than one billion of \BB events, which allows \babar\ and Belle collaborations to
perform precision measurements on \btosgam\ and \btodgam processes ~\cite{PDG}.
 Here I summarize the latest experimental achievements on the above inclusive processes. 


\section{Direct CP asymmetry in $B\rightarrow X_{s,d} \gamma$ }

The result presented\footnote{The branching fraction of $B\to X_s \gamma$ and its spectra shape from same analysis will be present in near future.}
 is based on a data sample of $\Pep\Pem \to \PgUc\to \BB$ 
collisions collected with the \babar\ detector at the PEP-II
asymmetric-energy \epem collider. The on-resonance integrated
luminosity is 347 $fb^{-1}$ and 36 $fb^{-1}$ of off-resonance data, taken 40 \mev
below the \PgUc resonance energy, are used to estimate the continuum
background ($\Pq\Paq:q=\Pqu\Pqd\Pqs\Pqc,\Pgtp\Pgtm$).
 
\begin{figure}[htb]
  \includegraphics[width=0.5\textwidth]{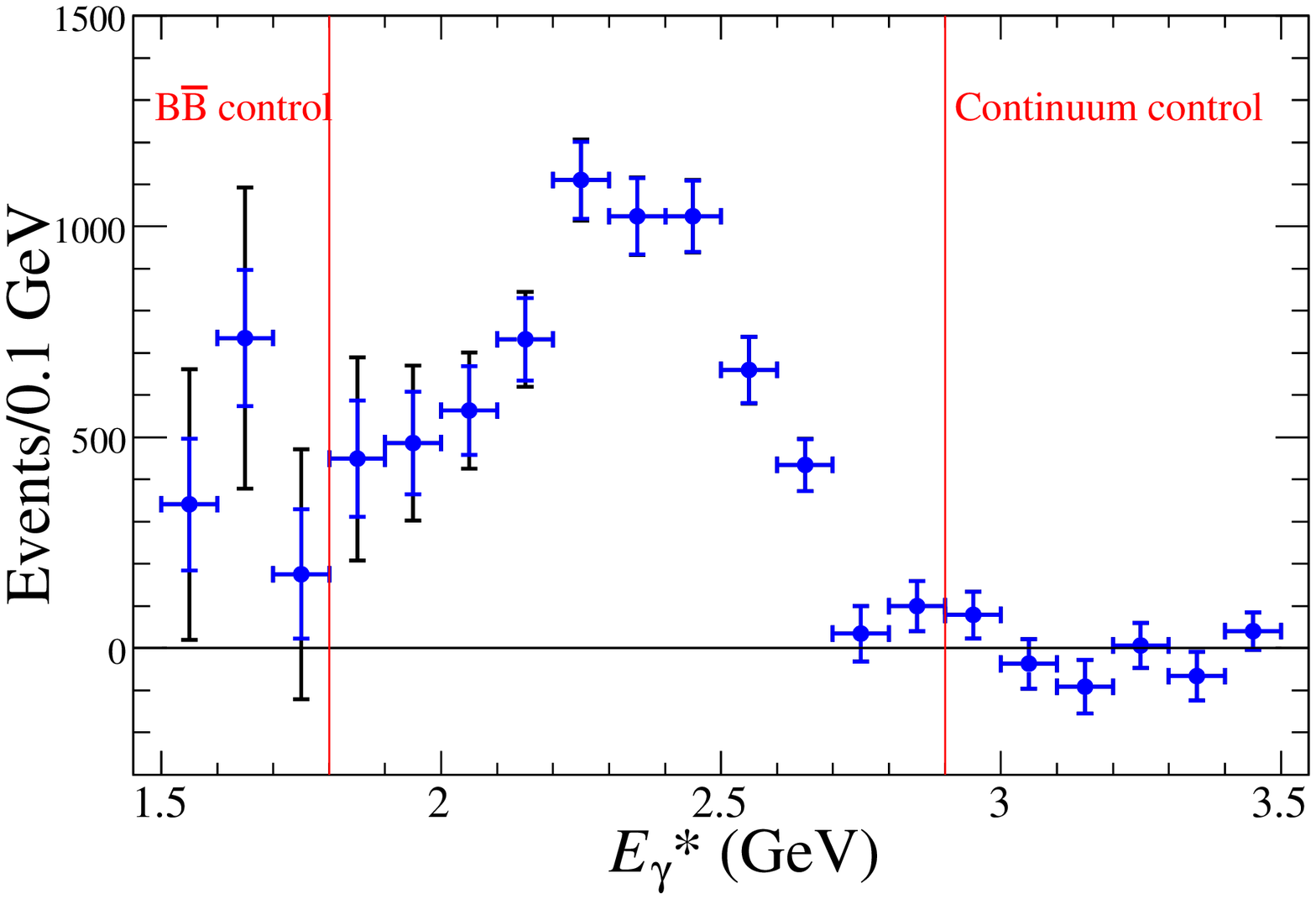}   
  \includegraphics[width=0.5\textwidth]{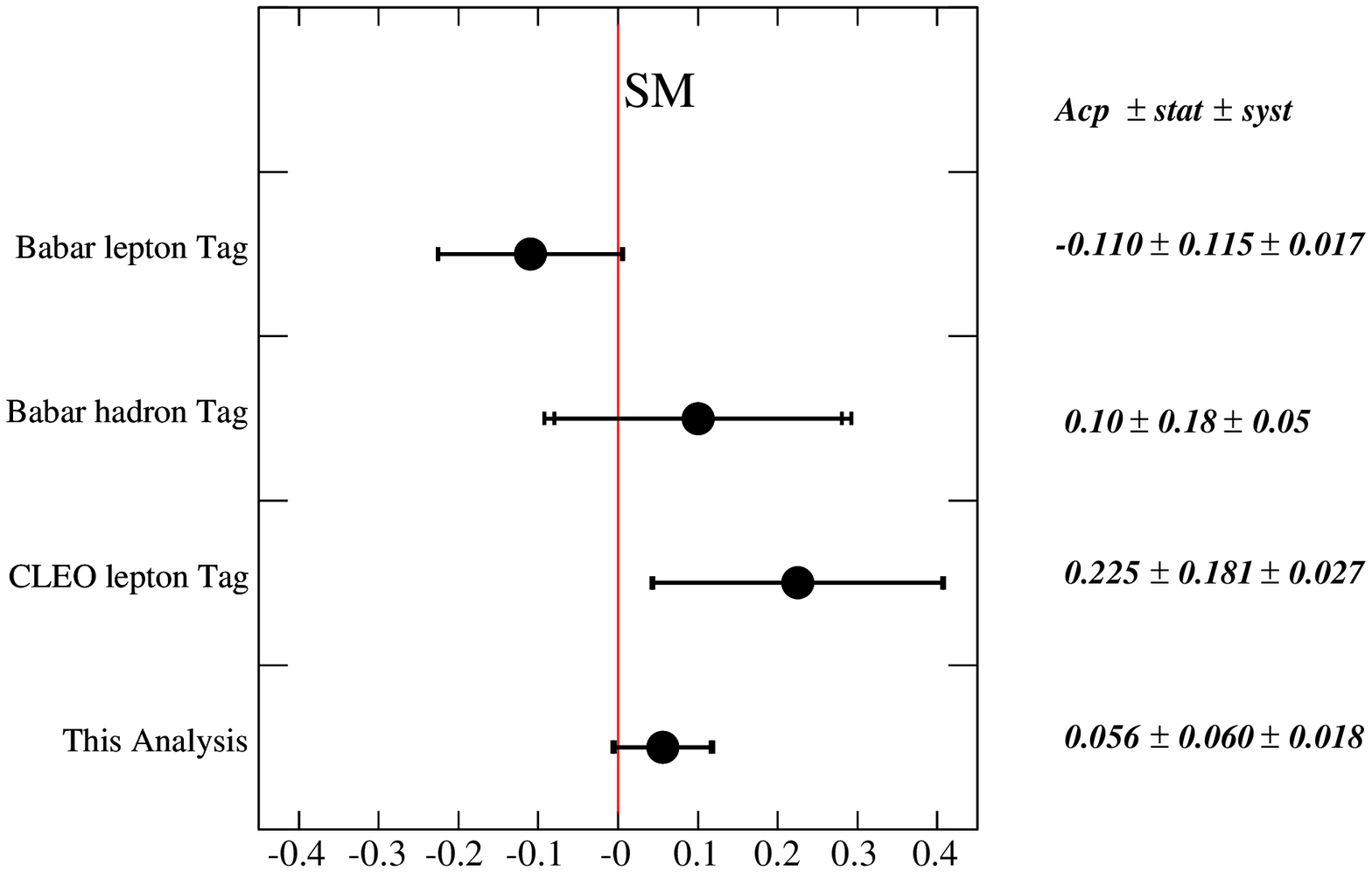}
\vspace{-0.2in}
\caption{Left: The photon spectrum in $347fb^{-1}$ of data after background subtraction.
 The inner error bars are statistical only, while the outer include both
 statistical and systematic errors in quadrature; 
Right: 
Measurements of \acp(\bxsdg), with statistical and systematic
   errors.  The three published results, top to
   bottom, are from references~\cite{BsgAcp}. }
\label{fig:egamma_dataminusbackgroundAndAcp} 
\end{figure}

 The analysis begins by requiring a high-energy photon, characteristic of $B\to X_s\gamma$ decays, while photons from $\pi^0$ and $\eta$ are vetoed. The background from continuum events  is significantly suppressed by charged lepton tagging and by exploiting the more jet-like topology of the $q\overline{q}$ or $\tau^+\tau^-$ events compared to the isotropic $B\overline{B}$ decays. The remaining continuum backgrounds are estimated with off-resonance data.
  The non signal \BB background arises predominantly from \Pgpz,\Pgh decay  but also from decays of other light mesons, mis-reconstructed electrons and hadrons, which are estimated using Monte Carlo simulation and corrected the data and MC difference using appropriate control samples.
 Figure~~\ref{fig:egamma_dataminusbackgroundAndAcp} shows the observed photo spectrum after subtracting off-resonance data and the corrected \BB backgrounds. 
Two prior selected control regions, \BB control ($1.53 < \egcms < 1.8 \gev$) and Continuum control ($2.9 < \egcms < 3.5 \gev $), are used to validate the background estimation. 
In the \BB control region we find $ 1252\pm 272 (stat.) \pm 841 (syst.)$ events,
dominated by \BB background with a small signal contribution component (~200-400 events depending on models); 
the continuum region yields s $ -100 \pm 138 (stat.)$ events, consistent with zero which showing good estimation of off-resonance subtraction.


The direct CP asymmetry, \acp(\bxsdg) 
is measured by dividing the signal sample into $\PB$ and $\PaB$ decays 
according to the charge of the lepton tag to measure 
$ \amcp(\bxsdg) = \frac{N^{+}-N^{-}}{N^{+}+N^{-}} $,
 where $N^{+(-)}$ are the positively (negatively) tagged signal yields. 
The asymmetry must be corrected for
the dilution due to the mistag fraction  $\omega$,
$\acp(\PB \to X_{s+d}\Pgg) = \frac{1}{1-2\omega}\amcp(\PB \to X_{s+d}\Pgg). $
the missing fraction $\omega$ is found to be $ 0.131 \pm 0.007$, from  $\PBz-\PaBz$ oscillation,
the fraction of events with  wrong-sign leptons from the \PB decay chain and the
similar fraction due to misidentification of hadrons as leptons.

The theoretical SM predictions for a near-zero asymmetry do not
require the entire spectrum to be measured.
To reduce the sensitivity to background, the signal region is restricted
to $2.1<\egcms<2.8\gev$. In this selected energy region,  the tagged signal yields are $N^{+}=2623 \pm 158 (stat.)$ and $N^{-}=2397 \pm 151(stat.)$
giving an asymmetry of  $\amcp(\bxsdg)=0.045 \pm 0.044\ .$
Finally the  $\amcp(\bxsdg)$ is corrected for mistagging and bias to give
$  \acp     =   0.056 \pm 0.060 (stat.) \pm 0.018 (syst.) $,
where the systematic error is mainly from non signal \BB background and the lepton tagging efficiency.
The result is consistent with no observed asymmetry, consistent with SM expectation and previous
measurements. A comparison of the result to published measurements is shown in
figure~\ref{fig:egamma_dataminusbackgroundAndAcp}.  The current measurement is the most precise
to date.


\section{Branching fraction and moments of $b\rightarrow s\gamma$ }

 Currently the most precise measurement of the inclusive $B\to X_s\gamma$ branching fractions has been done by Belle~\cite{Belle:Bsg2}.  The data consists of a sample of 605 $fb^{-1}$ taken on the $\Upsilon(4S)$ resonance. Another 68 $fb^{-1}$ sample was taken at an energy 60 MeV below the resonance. 

The signal spectrum is extracted by collecting all high energy photons, vetoing those originating from $\pi^0$ and $\eta$ decays to two photons. The non $B\overline{B}$ background, mainly $e^+e^-\rightarrow q\overline{q}$ ($q=u,d,s,c$) events, is subtracted using the off-resonance sample. The remaining background from $B\overline{B}$ are subtracted using Monte-Carlo simulated distributions normalized using data control samples. 

The analysis proceeds in two different streams, with lepton tag (LT) and without (MAIN). Two samples give 
similar sensitivity  to the signal while being largely statistically independent. After these selection criteria, $41.1 \times 10^5$ ( $24.6\times 10^4$) and $3.5\times 10^5$ ($0.9\times 10^4$) photon candidates survive in the MAIN (LT) stream of the on- and off-resonance data samples, respectively. 

The photon candidates from $B\overline{B}$ background is divided into six categories:(i) $\pi^0$; (ii)$\eta$; (iii) other real photons from decays of $\omega$, $\eta^{\prime}$ and $J/\psi$ mesons; (iv) mis-identified calorimeter clusters from $K^0_L$ and $\overline{n}$; (v) electrons misidentified as photons and; (vi) beam background. Each category is checked using appropriate control samples as described in Ref.\cite{Belle:Bsg1}.
Each background yield, scaled  by the described procedures, is subtracted from the data spectrum. The photon energy ranges 1.4-1.7 GeV and 2.8-4.0 GeV were chosen a prior as control regions to test the integrity of the background subtraction since in the low energy region the little signal expected is negligible with respect to the uncertainty on the background, and no signal is possible in the high energy region above the kinematic limit. The yield in the high energy region are $1245\pm4349$ and $292\pm410$ candidates in the MAIN and LT stream, respectively, while corresponding yields in the low energy region are $-1629\pm3071$ and $-745\pm623$, respectively.

 To obtain the true spectrum, a three-step unfolding procedure is used to correct the raw spectrum. 
The procedure does not distinguish between $B\to X_s \gamma$ and $B\to X_d \gamma$. Assuming the shape of the corresponding photon energy spectra are equivalent, the contribution of $B\to X_d \gamma$ is subtracted using the ratio
$R_{d/s}=(4.5\pm0.3)\%$. Boost corrections, obtained from MC simulation, are used to derive the measurements in the rest frame of the $B$ meson.
   The two streams, MAIN and LT, are combined taking the correlation into count. 

The measured branching fraction in the $B$-meson rest frame is $BF(B\to X_s\gamma)=(3.45\pm0.15\pm0.40)\times 10^{-4}$ for the photon energy range from 1.7 GeV to 2.8 GeV. The most accurate measurement is given in the photon energy range 2.0 GeV to 2.8 GeV,  $BF(B\to X_s\gamma)=(3.02\pm0.10\pm0.11)\times 10^{-4}$.   Here the errors are statistical and systematic, respectively. The measured branching fractions are in agreement with the latest theoretical calculation. The measured spectral moments can be used to reduce the uncertainty on $|V_{ub}|$~\cite{Simba,hfag}.


\section{$b\rightarrow d\gamma$ }

Here we present the first significant observation of 
the $\btodgam$ transition in the hadronic mass range $M(X_d)>1.0\gevcc$,
used in the determination
of $|V_{td}/V_{ts}|$ via the ratio of inclusive widths.
Inclusive $\BtoXsgam$ and $\BtoXdgam$ rates are extrapolated from the 
measurements of the partial decay rates of seven exclusive final states
 in the hadronic mass ranges  
$0.5<M(X_d)<1.0\gevcc$ and $1.0<M(X_d)<2.0\gevcc$.
Here the $X_d$ includes \pipi, \pipi\piz, \pipi \pip, \pipi \pipi, \pipi\pip\piz and $\pip\eta$; while the $X_{s}$ includes  \Kp\pim, \Kp \piz, \Kp \pipi, \Kp \pim \piz \Kp \pim \pipi, \Kp \pim \pip\piz and $\Kp\eta$. 
We combine these measurements and make a model-dependent extrapolation to higher 
hadronic mass to obtain an 
inclusive branching fraction (${\cal B})$ for $b\to (s,d)\gamma$.
These measurements use a sample of  $471 \times 10^{6}$ \BB\ pairs collected  by  the \babar\ experiment.

%

%
 
The signal yields in the data for the combination of all seven decay modes 
are determined from two-dimensional 
extended maximum likelihood fits to the $\de$ and $\mes$ distributions after all event selections, 
where $\de = E^*_B - E^*_{\rm beam}$,  $E^*_B$ is the energy of the $B$ meson 
candidate and $E^*_{\rm beam}$ is the beam energy, 
and  $\mes = \sqrt{ E^{*2}_{\rm beam}-{\vec{p}}_{B}^{\;*2}}$,  ${\vec{p}}_B^{\;*}$ is the momentum of the $B$ candidate. 
%
%
%
Table~\ref{tab:bfs} gives the signal yields, efficiencies 
and partial branching fractions.

\begin{table*}
\centering
\caption{\label{tab:bfs}
Signal yields ($N_S$), efficiencies ($\epsilon$). 
partial branching fractions ($BF$) and inclusive branching 
fractions ($\cal{B})$ for the measured decay modes. 
The first error is statistical the the second systematic (including 
error from extrapolation to missing decay modes, for the inclusive $\cal{B}$).}
{\small\tt
\begin{tabular}{ccccccc}
%
			& $M(X_s)0.4-1.0$	&$M(X_d) 0.4-1.0$	&$M(X_s) 1.0-2.0$	&$M(X_d)1.0-2.0$ (GeV/$c^2$)\   \\ \hline
$N_S$	 		&$804\pm 33$\ 		&  $35\pm 9$\		& $990\pm 42$\  	& $56\pm 14$\ 	    \\ 
$\epsilon$		&  4.5\%		& 3.1\%			& 1.6\%			& 1.9\%		    \\ 
$BF(\times 10^{-6})$ 	& $18.9\pm 0.8\pm 0.8$ 	& $1.2\pm 0.3\pm 0.1$ 	&  $65.7\pm 2.8\pm 5.9$ & $3.2\pm 0.8\pm 0.5$  \\ 
${\cal{B}}(\times 10^{-6})$ & $38.3\pm 1.6\pm 1.5$ & $1.3\pm 0.3\pm 0.1$ &$192\pm 80\pm 45$ & $7.9\pm 2.0\pm 3.3$  \\ \hline
$\frac{{\cal{B}}(b\to d\gamma)}{{\cal{B}}(b\to s\gamma)}$ & \multicolumn{2}{c}{$0.0033\pm 0.009\pm 0.003$}  & \multicolumn{2}{c}{-}  \\
\end{tabular} }
\end{table*}

To obtain inclusive ${\cal B}(\btosgam)$ and ${\cal B}(\btodgam)$ we need to 
correct the partial $\cal B$ values in Table~\ref{tab:bfs} 
for the fractions of missing final states. 
After correcting for the 50\% of missing decay modes with neutral kaons,
the low mass \BtoXsgam\ measurement is found to be consistent
with previous measurements of the rate for $B\to K^*\gamma$~\cite{PDG}.
For the low mass \BtoXdgam\ region, we correct for the small amount 
of non-reconstructed $\omega$ final states ($\omega \to \piz \gamma$ 
and others),  and find a partial
branching fraction consistent with previous measurements of 
$\BR(B\to(\rho,\omega)\gamma)$~\cite{PDG}. 
We assume that non-resonant decays do not contribute in this region. 

In the high mass region, 
the missing fractions depend on the fragmentation of the 
hadronic system and are expected to be different for $X_d$ and $X_s$.
We explore the uncertainty in the correction for missing modes by
considering several alternative models.
The resulting missing fractions vary by up
to 50\% relative to the nominal model.
We therefore independently vary final states with $\ge 5$ stable hadrons, or
with $\ge 2\pi^0$ or $\eta$ mesons,
by $\pm$50\%. 

 Combining the two mass regions, taking into account a
partial cancellation of the missing fraction errors in the ratio
of \btodgam\ to \btosgam, we find
${\cal B}(\btodgam) / {\cal B}(\btosgam) = 0.040 \pm 0.009(stat.) \pm 0.010(syst.) $
in the mass range $M(X)<2.0\gevcc$. 
For the unmeasured region $M(X) > 2.0$ \gevcc the differences between $\btosgam$
and $\btodgam$ are small and almost completely cancel in the ratio.

Conversion of the ratio of inclusive branching fractions to
the ratio $|V_{td}/V_{ts}|$ is done according to~\cite{AAG}, 
which requires $\bar{\rho}$ and $\bar{\eta}$ as input. 
However, since these are partially determined from previous 
measurements of $|V_{td}/V_{ts}|$, we instead re-express $\bar{\rho}$ and $\bar{\eta}$
 in terms of the independent CKM angle $\beta$. 
This procedure yields a value of 
$ |V_{td}/V_{ts}| = 0.199 \pm 0.022(stat.) \pm 0.024(syst.) \pm 0.002(th.)$ 
competitive with more model-dependent determinations
from the measurement of the exclusive modes
$B\to(\rho,\omega)\gamma$ and $B\to K^*\gamma$~\cite{bellerhog,babarrhog}.                     

\section{Summary}

Here I summarized the experiment progresses on the inclusive $\btosgam$ and $\btodgam$ after the last CKM workshop. Belle measured the inclusive branching fraction in the $B$-meson rest frame, $BF(B\to X_s\gamma)=(3.45\pm0.15\pm0.40)\times 10^{-4}$, for the photon energy range from 1.7 GeV to 2.8 GeV.
\babar\ presents the most precise direct CP asymmetry measurement to date, its preliminary result is consistent with SM prediction. \babar also measured the ratio of $b\rightarrow d\gamma$ over $b\rightarrow s\gamma$ using seven exclusive modes, providing the independent determination of $|V_{td}/V_{ts}|$.




\Acknowledgements
I am grateful to the wonderful works from \babar\ and Belle collaborations. Thanks also to the organizers of the CKM2010 for all efforts in making this venue successful.

\end{document}